\documentclass[lettersize,journal]{IEEEtran}
\usepackage{amsmath,amsfonts}
\usepackage{algorithmic}
\usepackage{algorithm}
\usepackage{array}
\usepackage[caption=false,font=normalsize,labelfont=sf,textfont=sf]{subfig}
\usepackage{textcomp}
\usepackage{stfloats}
\usepackage{url}
\usepackage{color}
\usepackage{booktabs}
\usepackage{verbatim}
\usepackage{graphicx}
\usepackage{multicol}
\usepackage{multirow}
\usepackage{cite}
\usepackage{amsmath}
\usepackage{float}
\usepackage{hyperref}
\hypersetup{
    colorlinks=true, 
    linkcolor=blue, 
    citecolor=blue,
    urlcolor=black  
}
\begin{document}

\title{Channel Knowledge Map Construction via Guided Flow Matching}

\author{Ziyu Huang, Yong Zeng,~\IEEEmembership{Fellow,~IEEE}, Shen Fu, Xiaoli Xu,~\IEEEmembership{Member,~IEEE,} and Hongyang Du,~\IEEEmembership{Member,~IEEE}

\thanks{Corresponding author: Xiaoli Xu.}
\thanks{Ziyu Huang, Shen Fu and Xiaoli Xu are with the National Mobile Communications
Research Laboratory, Southeast University, Nanjing 210096, China (e-mail:
\url{213220760@seu.edu.cn};  \url{sfu@seu.edu.cn};  \url{xiaolixu@seu.edu.cn}).}

\thanks{Yong Zeng is with the National Mobile Communications Research
Laboratory, Southeast University, Nanjing 210096, China, and also with the
Pervasive Communication Research Center, Purple Mountain Laboratories,
Nanjing 211111, China (e-mail: \url{yong_zeng@seu.edu.cn}).}

\thanks{Hongyang Du is with the Department of Electrical and Electronic Engineering, University of Hong Kong, Hong Kong SAR, China (e-mail:
\text{\url{duhy@eee.hku.hk}}).}
}
\markboth{}%
{Shell \MakeLowercase{\textit{et al.}}: A Sample Article Using IEEEtran.cls for IEEE Journals}

\IEEEpubid{}

\maketitle

\begin{abstract}
The efficient construction of accurate channel knowledge maps (CKMs) is crucial for unleashing the full potential of environment-aware wireless networks, yet it remains a difficult ill-posed problem due to the sparsity of available location-specific channel knowledge data. Although diffusion-based methods such as denoising diffusion probabilistic models (DDPMs) have been exploited for CKM construction, they rely on iterative stochastic sampling, rendering them too slow for real-time wireless applications. To bridge the gap between high fidelity and efficient CKM construction, this letter introduces a novel framework based on linear transport guided flow matching (LT-GFM). Deviating from the noise-removal paradigm of diffusion models, our approach models the CKM generation process as a deterministic ordinary differential equation (ODE) that follows linear optimal transport paths, thereby drastically reducing the number of required inference steps. We propose a unified architecture that is applicable to not only the conventional channel gain map (CGM) construction, but also the more challenging spatial correlation map (SCM) construction. To achieve physics-informed CKM constructions, we integrate environmental semantics (e.g., building masks) for edge recovery and enforce Hermitian symmetry for property of the SCM. Simulation results verify that LT-GFM achieves superior distributional fidelity with significantly lower Fréchet Inception Distance (FID) and accelerates inference speed by a factor of 25 compared to DDPMs.
\end{abstract}

\begin{IEEEkeywords}
Guided flow matching, channel knowledge map construction, environment-aware communication, diffusion models.
\end{IEEEkeywords}

\section{Introduction}
Acquiring location-specific prior channel knowledge is one of the most important aspects to realize the wireless digital twin\cite{wu2021digital}. To circumvent the prohibitive overhead of real-time channel estimation or environment sensing \cite{8226757}, the concept of channel knowledge map (CKM) has been proposed to enable the novel paradigm of environment-aware wireless networks\cite{9373011}, which makes use of historical  data to infer location-specific prior channel knowledge. CKMs can be categorized based on the type of channel knowledge they contain, such as channel gain, angle-of-arrival (AOA), path delay profile, and so on. This paper focuses on a holistic CKM that comprises two essential components: the channel gain map (CGM), which characterizes channel path loss and shadowing that are essential for power allocation and coverage prediction, and the spatial correlation map (SCM), which captures the multipath angular profiles that is useful for downstream tasks like multi-antenna beamforming\cite{adhikary2013joint}.

Constructing CKM from spatially sparse data poses a unique challenge. This task constitutes a severely ill-posed inverse problem. Existing solutions generally fall into three categories: numerical interpolation, deep learning-based regression, and generative modeling. Kriging interpolation is a classical example of interpolation. Though efficient, Kriging fails to capture complex non-linear relationships. Subsequently, deep regression methods (e.g., RadioUNet\cite{9354041}) have been extensively investigated. However, by optimizing pixel-wise mean squared error (MSE), they suffer from the ``regression-to-the-mean" phenomenon. For instance, for the SCMs construction task mentioned above, they tend to produce over-smoothed results that minimize average error but degrade the sharp eigen-structures required for precise spatial beamforming. To recover high-frequency details, generative models like GANs \cite{10130091} were introduced but remain notorious for training instability. While denoising diffusion probabilistic models (DDPMs) have recently achieved high generation quality\cite{10843401}\cite{fu2025ckmdiffgenerativediffusionmodel}, their application in wireless networks is hindered by excessive latency, as they require simulating a stochastic differential equation over hundreds of iterative steps.

To address these limitations, we propose an efficient and physically consistent CKM construction framework based on \textbf{guided flow matching} (GFM) with linear transport (LT) \cite{0f001e52f7724ff3a539cdef54b4a160}. Distinct from diffusion models that rely on stochastic noise removal, GFM learns a deterministic ordinary differential equation (ODE) that transports a simple prior distribution to the complex data distribution along a straight trajectory. This geometric simplicity allows the use of ODE solvers with large step sizes, significantly accelerating inference. More importantly, by modeling the probability flow rather than performing point-wise regression, our method preserves the statistical distribution of the channel features, ensuring low Fréchet inception distance (FID).

The main contributions of this letter are summarized as follows. First, we propose a novel flow matching-based framework for CKM construction. Unlike existing generative approaches that primarily focus on CGMs \cite{sun2025flowmatchingbasedactivelearning}, our framework establishes a unified approach applicable to both CGM and the more challenging SCM construction. Second, we design a multi-modal conditioning mechanism that incorporates environmental semantics to guide the generative flow for accurate CGM construction. In addition, for SCM construction, we impose a Hermitian symmetry constraint to ensure physical validity. Finally, simulation results demonstrate that our method achieves superior edge preservation and eigen-structure fidelity compared to the benchmarking DDPM-based method, all while accelerating the inference speed by 25 times.

\section{System Model and Problem Formulation}
\label{sec:system_model}

As shown in Fig. \ref{fig:background}, we consider a wireless communication system operating within a target region, denoted as $\mathcal{D}$. The region is discretized into a uniform 2D grid of size $H \times W$, where each pixel $(u, v)$ represents a potential user equipment (UE) location. The base station (BS), equipped with $N_t$ antennas, serves single-antenna UEs distributed across $\mathcal{D}$. We formulate the CKM construction problem as learning a conditional distribution $p(\mathbf{x}|\mathbf{c})$, where $\mathbf{x}$ denotes the high-fidelity channel knowledge whose dimension is to be specified based on the application,  and $\mathbf{c}$ denotes the available degraded observation. This general formulation includes two specific sub-problems as special cases:

\begin{figure}
    \centering
    \includegraphics[width=1.0\linewidth]{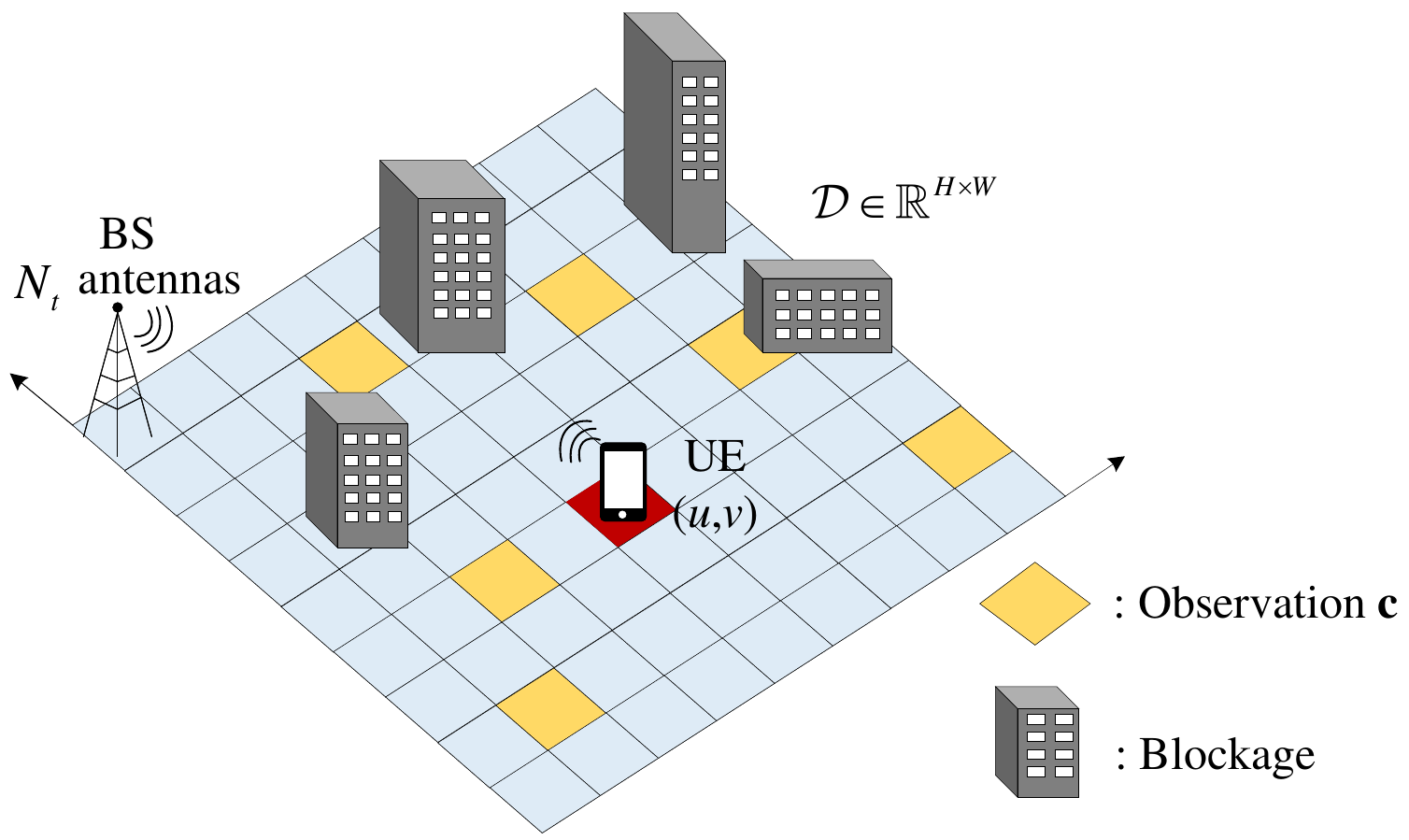}
    \caption{CKM constructions using environment semantics.}
    \label{fig:background}
\end{figure}

\begin{figure}
    \centering
    \includegraphics[width=1\linewidth]{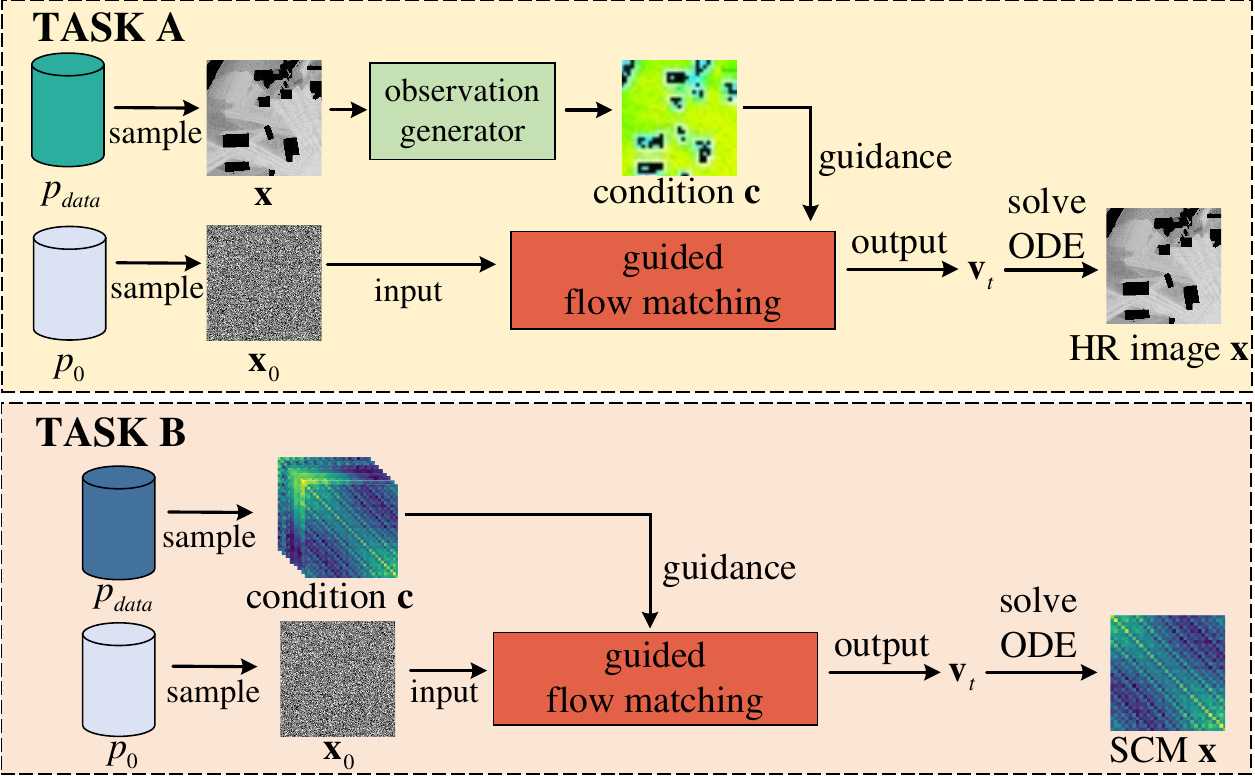}
    \caption{Framework for different tasks in CKM constructions.}
    \label{fig:placeholder}
\end{figure}
\textbf{A: CGM construction:} The goal is to recover the high-resolution CGM from coarse observations. Let $\mathbf{x}\in \mathbb{R}^{HW \times 1}$ represent the vectorized ground-truth CGM (in dB). The noisy and incomplete channel knowledge observed can be represented as: $\mathbf{y} = \mathbf{A}\mathbf{x} + \mathbf{n}$, where $\mathbf{y}\in\mathbb{R}^{hw\times1}$ denotes the observed channel gain data, $\mathbf{A}\in\{0,1\}^{hw\times HW}(hw\ll HW)$ denotes the downsampling matrix, and $\mathbf{n}$ is the additive Gaussian noise with mean $\mathbf{0}$ and covariance matrix $\sigma^2\mathbf{I}$. To incorporate environmental semantics, we apply both segments extraction algorithm and canny detection algorithm to introduce a building mask $\mathbf{m}$ and an edge map $\mathbf{e}$ based on the observation $\mathbf{y}$. We define $\mathbf{m}_{u,v} = \mathbb{I}((u,v) \notin \mathcal{B})$ to indicate valid outdoor regions, and $\mathbf{e}_{u,v} = \mathbb{I}((u,v) \in \partial \mathcal{B})$ to highlight diffraction edges, where $\mathbb{I}(\cdot)$ is the indicator function and $\mathcal{B}$ denotes the blockage. These inputs are concatenated to form the condition tensor $\mathbf{c} \in \mathbb{R}^{3 \times hw \times 1}$:
\begin{equation}
    \mathbf{c} = [\mathbf{y}, \mathbf{m}, \mathbf{e}].
    \label{eq:originalinput}
\end{equation}
Our framework aims to learn the conditional distribution $p(\mathbf{x}|\mathbf{c})$ to recover the high-resolution texture of $\mathbf{x}$ by guiding the flow with these geometric priors.

\textbf{B: SCM construction:} The goal of this task is to recover the missing SCM at a target location $(u,v)$ by leveraging the spatial consistency of wireless channels. To formulate this problem, we first establish the forward physical model. The ground-truth SCM field can be modeled as a spatially continuous function governed by the local scattering environment. For any location $\mathbf{p} = (u, v)$, the SCM $\mathbf{R}(\mathbf{p})$ is defined by:
\begin{equation}
    \mathbf{R}(\mathbf{p}) = \mathbb{E}\left[ \mathbf{h}(\mathbf{p}) \mathbf{h}^H(\mathbf{p}) \right] ,
    \label{eq:scm_physics}
\end{equation}
where $\mathbf{h}(\mathbf{p})$ is the channel vector. Due to the continuity of electromagnetic propagation, the SCM at $\mathbf{R}(\mathbf{p})$ is statistically dependent on the SCMs of its surrounding locations. This dependency can be represented as $\mathbf{R}(\mathbf{p}) \approx \Psi \left( \{ \mathbf{R}(\mathbf{q}) \mid \mathbf{q} \in \mathcal{N}_{\mathbf{p}} \} \right)$, where $\mathcal{N}_{\mathbf{p}}$ denotes the set of spatial neighbors around $\mathbf{p}$, and $\Psi(\cdot)$ represents the sophisticated mapping induced by the physical environment.

Based on the forward model, we define the SCM construction as a learning problem to approximate $\Psi$. We define the target data $\mathbf{x} \in \mathbb{R}^{2 \times N_t \times N_t}$ as the real-valued tensor decomposition of the ground-truth  SCM, i.e., $\mathbf{x} = [\Re\{\mathbf{R}(\mathbf{p})\}; \Im\{\mathbf{R}(\mathbf{p})\}]$. For the observation, we assume a sparse sampling scenario where the target SCM is unknown, but a set of $K$ neighboring tensors $\mathbf{y} = \{ \mathbf{x}_1, \cdots, \mathbf{x}_K \}$ is available. The condition tensor $\mathbf{c}$ is constructed by stacking these neighbors along the channel dimension:
\begin{equation}
    \mathbf{c} = [\mathbf{x}_1, \cdots, \mathbf{x}_K] \in \mathbb{R}^{2K \times N_t \times N_t}.
    \label{eq:condition_construction}
\end{equation}
Our goal is to learn the conditional distribution $p(\mathbf{x}|\mathbf{c})$ to construct the center SCM $\mathbf{x}$ given the spatial context $\mathbf{c}$. The  framework of our tasks are illustrated in Fig. \ref{fig:placeholder}.

\section{Guided Flow Matching Framework for CKM Construction }
\label{sec:method}

\subsection{Generative CKM Flow via Optimal Transport}

Constructing a complete CKM from sparse observations is a severely ill-posed inverse problem, implying that infinite potential CKMs $\mathbf{x}$ could fit the limited observation $\mathbf{y}$. Instead of deterministically estimating a single solution, which often leads to blurry averages, we formulate the construction as learning the conditional probability distribution of physically valid channels. In order to define the process, we introduce three key distributions. The \textbf{target distribution} $p_{data}(\mathbf{x})$ represents the ground-truth manifold of valid CKM for $\mathbf{x}$. The \textbf{prior distribution} $p_0$ serves as our starting point, defined as a standard Gaussian distribution $p_0(\mathbf{x}) \sim \mathcal{N}(\mathbf{0}, \mathbf{I})$ with zero information. The \textbf{generated distribution} $p_1$ is the output of our model. Our fundamental goal is to transform the prior distribution $p_0$ into $p_1$ such that $p_1$ well approximates the true physical distribution $p_{data}$.

To achieve the transformation mentioned above, we adopt the flow matching paradigm. Unlike diffusion models that rely on stochastic Langevin dynamics \cite{10081412}, our framework models the CKM generation as a deterministic flow governed by an ODE:
\begin{equation}
    \begin{cases}
        \frac{d\mathbf{x}_t}{dt} = \mathbf{v}_{\theta}(\mathbf{x}_t, t, \mathbf{c}), \quad t \in [0, 1],\\
        \mathbf{x_0}\sim p_0(\mathbf{x}),\mathbf{x}_1\sim p_1(\mathbf{x})
    \end{cases}
    \label{eq:ode}
\end{equation}
where $\mathbf{x}_t$ denotes the intermediate CKM state at time $t$, and $\mathbf{v}_{\theta}$ is a neural network parameterizing the velocity field. Starting from pure noise at $t=0$, we integrate this ODE to progressively ``denoise" the state, evolving it into a valid CKM at $t=1$.

While numerous paths exist to transfer  $p_0$ to $p_{data}$, we select the LT path for computational efficiency. Geometrically, the LT path \textbf{directly} connects a noise sample $\mathbf{x}_0$ and a data sample $\mathbf{x}_1$ via a straight line trajectory defined as $\mathbf{x}_t = (1-t)\mathbf{x}_0 + t\mathbf{x}_1$. This linear interpolation implies that the target velocity field $\mathbf{u}_t$ is a constant vector pointing directly from noise to data: $\mathbf{u}_t = \mathbf{x}_1 - \mathbf{x}_0$. Consequently, network training simplifies to a regression problem where we minimize the difference between the predicted velocity $\mathbf{v}_{\theta}$ and the linear direction:
\begin{equation}
    \mathcal{L}_{GFM}(\theta) = \mathbb{E}_{t, \mathbf{x}_0, \mathbf{x}_1,\mathbf{c}} \left[ \| \mathbf{v}_{\theta}(\mathbf{x}_t, t, \mathbf{c}) - (\mathbf{x}_1 - \mathbf{x}_0) \|^2 \right].
    \label{eq:loss}
\end{equation}
Minimizing Eq. \eqref{eq:loss} naturally enables us to convert $\mathbf{x}_0$ to $\mathbf{x}_1$ smoothly.

\subsection{Training and Inferring}
\begin{figure}[ht]
    \centering
    \includegraphics[width=1\linewidth]{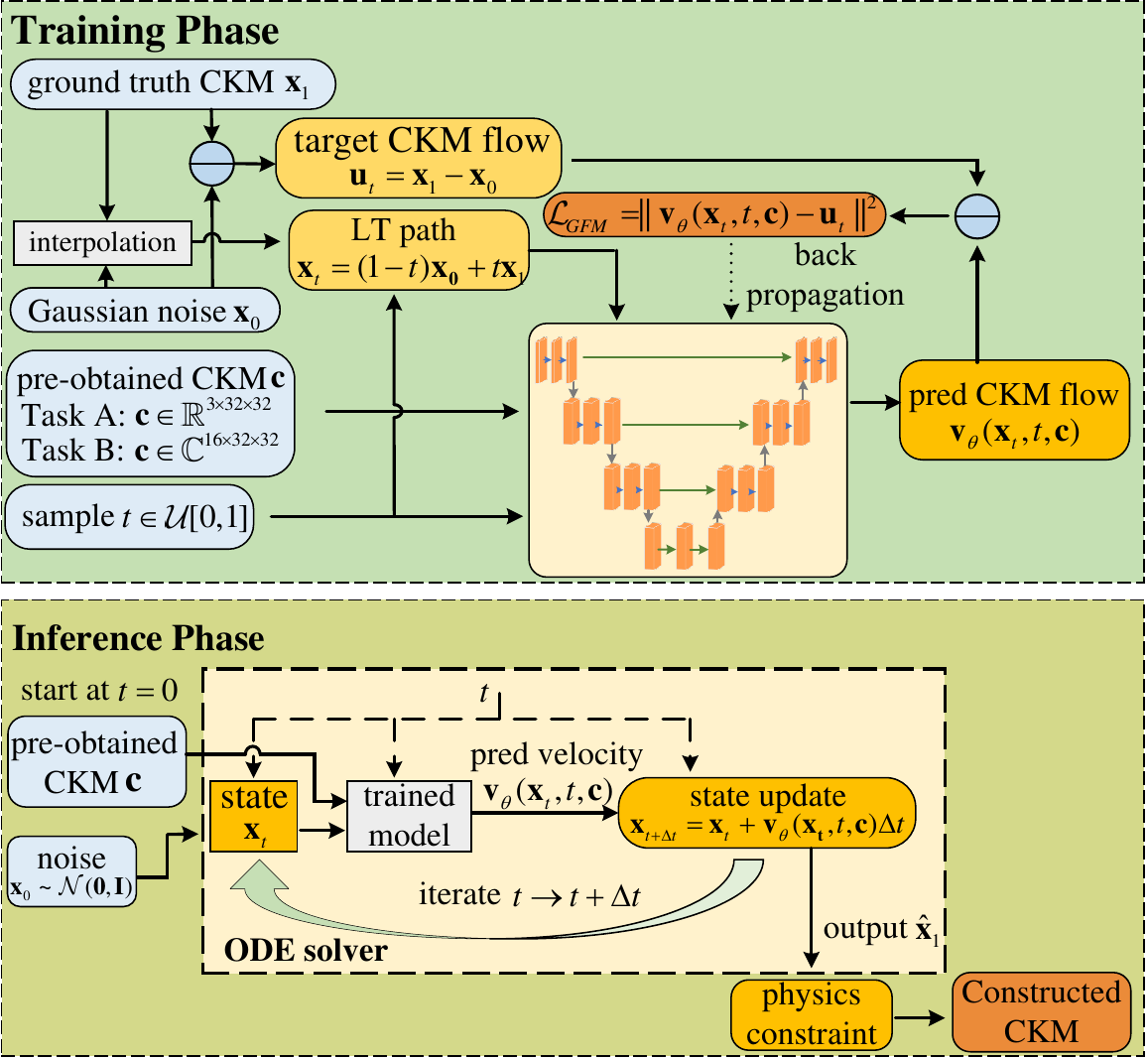}
    \caption{The block diagram of the training and inference phases of the proposed CKM construction framework.}
    \label{fig:system_overview}
\end{figure}

As illustrated in Fig. \ref{fig:system_overview}, the proposed framework operates in two phases: learning the conditional velocity field via regression  and generating samples via numerical integration, which are also known as training phase and inference phase.

\subsubsection{Training Phase}
The training objective is to optimize the network parameters $\theta$ to approximate the LT path. The procedures are outlined in Algorithm \ref{alg:cfm_training}. Our training utilizes dataset pairs $(\mathbf{c}, \mathbf{x}_1)$ derived from the ground-truth samples.Specifically, to adapt the low-resolution observations to the U-Net architecture in Task A, we employ an explicit condition alignment strategy. The raw observation $\mathbf{y}$ is pre-upsampled to the target domain $H \times W$ using a bicubic operator $\mathcal{U}(\cdot)$ and concatenated with the building mask $\mathbf{m}$ and edge map $\mathbf{e}$, yielding the condition tensor $\mathbf{c} = [\mathcal{U}(\mathbf{y}), \mathcal{U}(\mathbf{m}), \mathcal{U}(\mathbf{e})]$. For Task B, $\mathbf{c}$ is formed by stacking neighboring SCMs. 

The model learns to regress a constant velocity field. By sampling a time step $t \sim \mathbb{U}[0, 1]$ and constructing the intermediate state $\mathbf{x}_t = (1-t)\mathbf{x}_0 + t\mathbf{x}_1$, we enforce the probability mass to move along straight lines. The network parameters are updated by minimizing the mean squared error (MSE) between the predicted velocity $\mathbf{v}_{\text{pred}}$ and the target direction $\mathbf{u}_{\text{target}} = \mathbf{x}_1 - \mathbf{x}_0$, ensuring the learned flow points directly from the noise prior to the data manifold.

\subsubsection{Inference Phase}
During inference, the model constructs high-fidelity CKM from pure noise $\mathbf{x}_0 \sim \mathcal{N}(\mathbf{0}, \mathbf{I})$ by solving the learned ODE, as detailed in Algorithm \ref{alg:cfm_inference}. We employ the Euler discretization method to evolve the state from $t=0$ to $t=1$. Throughout this process, the condition $\mathbf{c}$ remains constant, acting as a navigational guide for the flow trajectory. Upon completion of the integration, task-specific post-processing is applied. 
For Task B, the solver output $\hat{\mathbf{x}}$ represents the real and imaginary parts of the matrix. To enforce the structural consistency of the constructed channel covariance, we execute a symmetrization step: $\mathbf{R}_{final}=\frac{1}{2}(\mathbf{R}+\mathbf{R}^{H})$. This operation strictly imposes the Hermitian symmetry property, which is a fundamental prerequisite for valid SCMs and essential for stable downstream beamforming.

\vspace{-2pt}
\begin{algorithm}[h]
\caption{Training Procedure for CKM construction via Guided Flow Matching}
\label{alg:cfm_training}
\begin{algorithmic}[1]
\STATE \textbf{Input:} Pre-normalized dataset $\mathcal{S  } = \{(\mathbf{c}, \mathbf{x}_1)\}$ containing condition tensors and ground-truth targets.
\STATE \textbf{Hyperparameters:} Batch size $B$, Learning rate $\eta$, Epoch $N$.
\STATE \textbf{Initialize:} Neural network parameters $\theta$ (DiffUNet), Optimizer.

\WHILE{$i<N$}

    \STATE Sample a mini-batch of pairs$(\mathbf{c},\mathbf{x}_1)\sim\mathcal{D}$, initial Gaussian noise $\mathbf{x}_0 \sim \mathcal{N}(\mathbf{0}, \mathbf{I})$, random time steps $t \sim \mathcal{U}[0, 1]$.

    \STATE Construct linear transport path:
            $\mathbf{x}_t = (1 - t) \mathbf{x}_0 + t \mathbf{x}_1$

    \STATE Compute conditional target velocity:
        $\mathbf{u}_{\text{target}} = \mathbf{x}_1 - \mathbf{x}_0$

    \STATE Predict by neural network:
        $\mathbf{v}_{\text{pred}} = \mathbf{v}_\theta(\mathbf{x}_t, t, \mathbf{c})$

    \STATE Compute loss:
    
        $\mathcal{L}(\theta) 
        = \frac{1}{B} \sum \left\| \mathbf{v}_{\text{pred}} - \mathbf{u}_{\text{target}} \right\|^2$

    \STATE Update network parameters:
        $\theta \leftarrow \theta - \eta \nabla_\theta \mathcal{L}(\theta)$

\ENDWHILE

\STATE \textbf{Output:} Optimized model parameters $\theta^*$
\end{algorithmic}
\end{algorithm}

\begin{algorithm}[ht]
\caption{Inference Procedure for CKM construction via ODE Sampling}
\label{alg:cfm_inference}
\begin{algorithmic}[1]
\STATE \textbf{Input:} Condition tensor $\mathbf{c}$ , Pre-trained FlowUNet parameters $\theta$, Number of integration steps $N$, Channel-wise statistics $(\boldsymbol{\mu}, \boldsymbol{\sigma})$ pre-computed from the training set.
\STATE \textbf{Initialize:} Sample initial Gaussian noise $\mathbf{x}_0 \sim \mathcal{N}(\mathbf{0}, \mathbf{I})$.
\STATE \textbf{Preprocessing:} $\tilde{\mathbf{c}} \leftarrow (\mathbf{c} - \boldsymbol{\mu}) / \boldsymbol{\sigma}$.
\STATE Set time step size $\Delta t = 1/N$.

\WHILE{$i<N$}
    \STATE Current time embedding $t \leftarrow i/N$.
    
    \STATE Predict via neural network:
        $\mathbf{v}_{\text{pred}} = \mathbf{v}_\theta(\mathbf{x}_t, t, \tilde{\mathbf{c}})$
        
    \STATE Evolve state:
        $\mathbf{x}_{t+\Delta t} = \mathbf{x}_t + \mathbf{v}_{\text{pred}} \cdot \Delta t$
\ENDWHILE

\STATE \textbf{Post-processing: }$\hat{\mathbf{x}} \leftarrow \mathbf{x}_1 \cdot \boldsymbol{\sigma} + \boldsymbol{\mu}$.

\IF{Task B}
    \STATE Construct $\mathbf{R}$ from real \& imaginary channels of $\hat{\mathbf{x}}$.
    \STATE Set physical constraint):
        $\mathbf{R}_{\text{final}} = \frac{1}{2} (\mathbf{R} + \mathbf{R}^H)$
    \STATE \textbf{Output:} $\mathbf{R}_{\text{final}}$
\ELSE
    \STATE \textbf{Output:} Super-resolved CGM $\hat{\mathbf{x}}$
\ENDIF
\end{algorithmic}
\end{algorithm}

\section{Simulation Results}
\label{sec:results}

\subsection{Simulation Setup}
\label{subsec:simulation_setup}
Our experiments utilize the \textit{CKMImageNet} dataset \cite{11184538}, generated via Wireless InSite simulation software.
For \textbf{Task A}, the raw CGMs ($128 \times 128$) are normalized to grayscale images. To simulate realistic degradation, the high-resolution maps, in which he value of each pixel is between 0 and 255, are down-sampled by a factor of 4 and contaminated with Gaussian noise at pixel level ($\mathbf{n}\sim\mathcal{N}(\mathbf{0}, 30^2\mathbf{I})$).
For \textbf{Task B}, we consider a BS with $N_t=32$ antennas, and $K=8$, indicating 8 surrounding SCMs are given to predict SCM at target locations.

\subsection{Performance Metrics}
\label{subsec:metrics}
We employ a diverse set of metrics summarized in Table \ref{tab:metrics_summary_simple}. Ordinary metrics including pixel-level MSE, peak signal-to-noise ratio (PSNR), and structural similarity index measure (SSIM) are used for pixel-level and perceptual evaluation.
To assess the distributional fidelity and physical consistency, we specifically adopt:

\subsubsection{\textbf{Fréchet inception distance (FID)}}
FID measures the Wasserstein-2 distance between feature distributions of real and generated data. Lower FID indicates higher statistical realism:
\begin{equation}
\text{FID} = \|\boldsymbol{\mu}_r - \boldsymbol{\mu}_g\|_2^2 + \text{Tr}\left( \mathbf{\Sigma}_r + \mathbf{\Sigma}_g - 2(\mathbf{\Sigma}_r \mathbf{\Sigma}_g)^{1/2} \right),
\end{equation}
where $(\boldsymbol{\mu}_r, \mathbf{\Sigma}_r)$ and $(\boldsymbol{\mu}_g, \mathbf{\Sigma}_g)$ denote the empirical sample mean vectors and covariance matrices calculated from the deep feature representations of the real dataset (ground truth) and the generated dataset, respectively \cite{LU20081044}.

\subsubsection{\textbf{Matrix similarity index (MSI)}}
For Task B, MSI evaluates the preservation of the SCM's eigen-space structure:
\begin{equation}
\text{MSI}(\mathbf{R}, \hat{\mathbf{R}}) = \frac{1}{N} \sum_{n=1}^{N} \frac{\left| \text{Tr}(\mathbf{R}_n \hat{\mathbf{R}}_n^H) \right|}{\|\mathbf{R}_n\|_F \|\hat{\mathbf{R}}_n\|_F},
\end{equation}
where $N$ denotes the total number of evaluation samples. $\mathbf{R}_n$ and $\hat{\mathbf{R}}_n$ represent the ground-truth and constructed SCM of the $n$-th sample, respectively. The term $\|\cdot\|_F$ denotes the Frobenius norm, and $(\cdot)^H$ represents the conjugate transpose.

\begin{table}[htbp]
\centering
\caption{Applicability of Performance Metrics}
\label{tab:metrics_summary_simple}
\begin{tabular}{cccc}
\toprule
\textbf{Category} & \textbf{Metric} & \textbf{Task A} & \textbf{Task B} \\
\midrule
Pixel-level & MSE, RMSE, NMSE & \checkmark & \checkmark \\
Perceptual & PSNR, SSIM, FID & \checkmark & \text{FID} \\
Physical & MSI & \text{--} & \checkmark \\
\bottomrule
\end{tabular}
\end{table}

\subsection{Numerical Results}
\label{subsec:results}

We compare the  proposed LT-GFM framework against baselines including interpolation methods (Bicubic, Bilinear, KNN), UNet-based regression, and DDPM.

\begin{figure*}[htbp]
    \centering
    \includegraphics[width=1\linewidth]{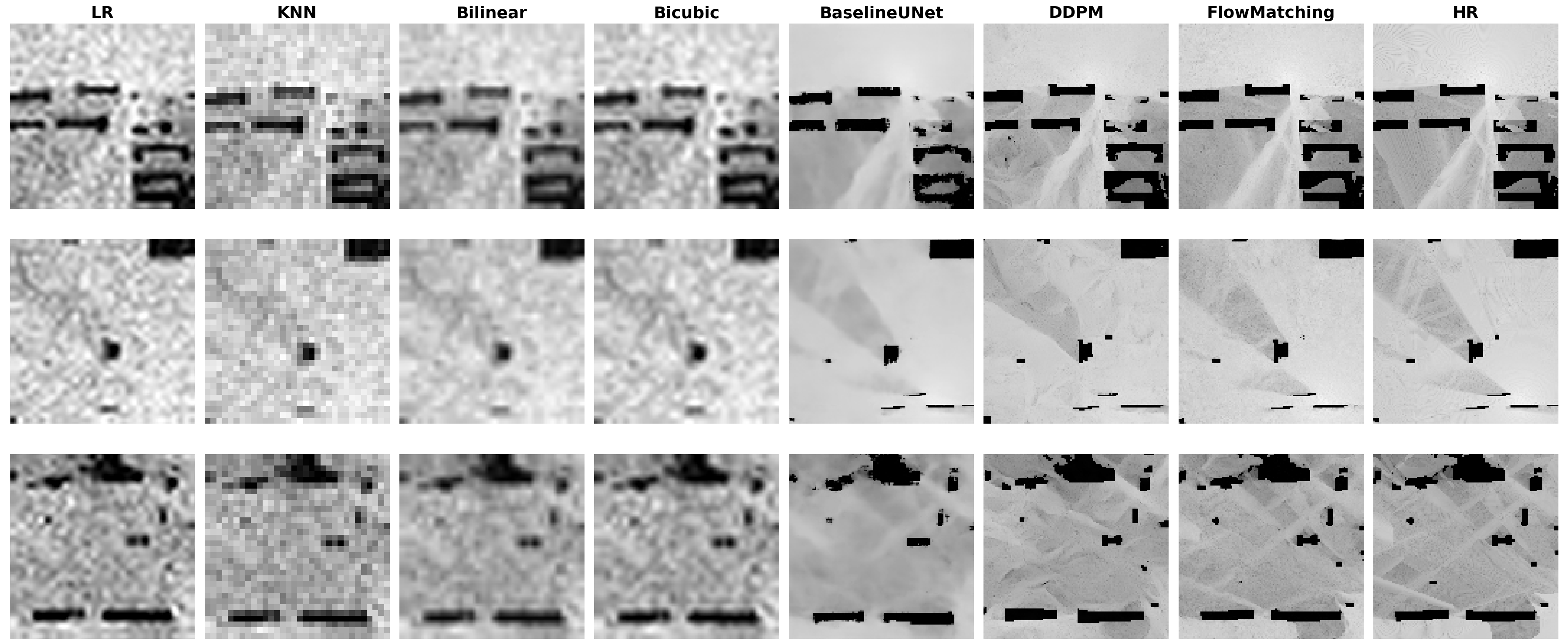}
    \caption{The comparisons of constructed CGM on different methods}
    \label{fig:Ex1}
\end{figure*}

\textbf{1) Task A:} As shown in Table \ref{tab:task_a_results}, our method achieves the lowest NMSE and the highest SSIM (0.8341), significantly outperforming the regression baseline (0.6624). Fig. \ref{fig:Ex1} illustrates a set of visualized figures using different methods. In terms of generative quality, the FID score is reduced by $43\%$ compared to DDPM, indicating that the deterministic flow generates sharper and more realistic textures (e.g., building shadows). Crucially, this performance gain is achieved with superior efficiency: our method reduces the inference time by approximately $25\times$ compared to DDPM (417 ms vs. 10428 ms). The speedup verifies that the straight LT probability path effectively overcomes the computational bottleneck of traditional stochastic sampling.

\begin{table}[htbp]
\centering
\caption{Numerical Comparison for Task A (CGM construction)}
\label{tab:task_a_results}
\setlength{\tabcolsep}{0pt} 
\begin{tabular*}{\linewidth}{@{\extracolsep{\fill}}cccccc}
\toprule
\textbf{Method} & \textbf{NMSE}$\downarrow$ & \textbf{PSNR}$\uparrow$ & \textbf{SSIM}$\uparrow$ & \textbf{FID}$\downarrow$ & \textbf{Time(ms)}\\
\midrule
KNN & 0.0735 & 16.55 & 0.2935 & 300.74 & - \\
Bilinear & 0.0612 & 17.44 & 0.3862 & 263.91 & - \\
Bicubic & 0.0641 & 17.15 & 0.3410 & 248.05 & - \\
BaselineUNet & 0.0558 & 17.93 & 0.6624 & 184.58 & 149 \\
DDPM & 0.0368 & 20.53 & 0.6392 & 107.69 & 10428 \\
\textbf{Proposed} & \textbf{0.0021} & \textbf{32.49} & \textbf{0.8341} & \textbf{61.02} & \textbf{417} \\
\bottomrule
\end{tabular*}
\end{table}

\textbf{2) Task B:} Table \ref{tab:task_b_results} presents the construction performance for SCMs. While UNet yields a low MSE, its high FID (23.78) suggests a loss of distributional characteristics. In contrast, our method achieves a near-perfect MSI of 0.90 and a significantly lower FID of 1.40. This confirms that the framework preserves the eigen-structure which will influence significantly for downstream task like beamforming, avoiding the ``over-smoothing'' effect. Specifically, while regression-based baselines tend to blur fine-grained spatial correlation features, our generative flow accurately synthesizes the high-frequency details consistent with the scattering environment.

\begin{table}[htbp]
\centering
\caption{Numerical Comparison for Task B (SCM construction)}
\label{tab:task_b_results}
\setlength{\tabcolsep}{0pt}
\begin{tabular*}{\linewidth}{@{\extracolsep{\fill}}ccccc}
\toprule
\textbf{Method} & \textbf{RMSE}$\downarrow$ & \textbf{NMSE}$\downarrow$ & \textbf{MSI}$\uparrow$ & \textbf{FID}$\downarrow$ \\
\midrule
KNN & 0.012 & 0.35 & 0.77 & 63.71\\
UNet & 0.004 & 0.17 & 0.91 & 23.78 \\
\textbf{Proposed} & \textbf{0.006} & \textbf{0.22} & \textbf{0.90} & \textbf{1.40}\\
\bottomrule
\end{tabular*}
\end{table}

\subsection{Ablation Study}
\label{subsec:ablation}
We investigate the impact of the number of ODE sampling steps $N$ on construction quality and latency. As illustrated in Table. \ref{tab:ablation_steps}, the performance (SSIM) saturates around $N=10$. Increasing $N$ further yields marginal gains but linearly increases computational cost. Thus, $N=10$ strikes a good balance, achieving high fidelity with an inference time of only 417 ms.
Furthermore, conditioning analysis shows that removing geometric priors (mask \& edge) causes the SSIM to drop from 0.858 to 0.564, validating the necessity of environmental semantics.

\begin{table}[htbp]
  \centering
  \caption{Ablation Study on Sampling Steps}
  \label{tab:ablation_steps}
  \begin{tabular}{ccc}
    \toprule
    \textbf{Steps ($N$)} & \textbf{SSIM} & \textbf{Time (ms)} \\
    \midrule
    1   & 0.3674 & 77.75   \\
    2   & 0.5345 & 150.53  \\
    4   & 0.7194 & 299.35  \\
    8   & 0.8168 & 598.08  \\
    10  & 0.8356 & 745.08  \\
    20  & 0.8562 & 1489.94 \\
    50  & 0.8569 & 3726.70 \\
    \bottomrule
  \end{tabular}
\end{table}

\vspace{-10pt}
\section{Conclusion}
In this letter, we proposed a novel generative framework for CKM construction based on linear transport guided flow matching. By modeling the generative process as a deterministic ODE with straight trajectories, our approach effectively overcomes the computational bottleneck of traditional diffusion models. We incorporated environmental semantics and physical constraints into the flow learning process, enabling the accurate recovery of both large-scale shadowing effects and small-scale spatial correlation structures. Numerical results confirm that the our method achieves superior construction accuracy, characterized by preserved eigen-structures and realistic textures. Most critically, it accelerates the inference speed by an order of magnitude ($25\times$) compared to DDPM. This combination of high fidelity and low latency validates the potential of flow matching as a practical solution for real-time CKM constructions in future 6G networks.

\bibliographystyle{IEEEtran} 
\bibliography{IEEEabrv, reference}

@article{11184538,
  author   = {Wu, Zijian and Wu, Di and Fu, Shen and Qiu, Yuelong and Zeng, Yong},
  journal  = IEEE_J_COM,
  title    = {{CKMImageNet}: A Dataset for {AI}-Based Channel Knowledge Map Towards Environment-Aware Communication and Sensing},
  year     = {2025},
  volume   = {},
  number   = {},
  pages    = {1-1},
  doi      = {10.1109/TCOMM.2025.3615778}
}

@article{9373011,
  author   = {Zeng, Yong and Xu, Xiaoli},
  journal  = IEEE_M_WC,
  title    = {Toward Environment-Aware {6G} Communications via Channel Knowledge Map},
  year     = {2021},
  month    = jun,
  volume   = {28},
  number   = {3},
  pages    = {84-91},
  doi      = {10.1109/MWC.001.2000327}
}

@inproceedings{0f001e52f7724ff3a539cdef54b4a160,
  title     = {Flow Matching for Generative Modeling},
  author    = {Lipman, Yaron and Chen, Ricky T. Q. and Ben-Hamu, Heli and Nickel, Maximilian and Le, Matt},
  booktitle = {Proc. Int. Conf. Learn. Repr. (ICLR)},
  year      = {2023},
  month     = may
}

@article{fu2025ckmdiffgenerativediffusionmodel,
      title={CKMDiff: A Generative Diffusion Model for CKM Construction via Inverse Problems with Learned Priors}, 
      author={Shen Fu and Yong Zeng and Zijian Wu and Di Wu and Shi Jin and Cheng-Xiang Wang and Xiqi Gao},
      year={2025},
      eprint={2504.17323},
     journal ={arXiv},
}

@article{sun2025flowmatchingbasedactivelearning,
      title={Flow Matching-Based Active Learning for Radio Map Construction with Low-Altitude UAVs}, 
      author={Hao Sun and Shicong Liu and Xianghao Yu and Ying Sun},
      year={2025},
      eprint={2509.13822},
      journal={arXiv},
      primaryClass={eess.SP}, 
}

@article{LU20081044,
title = {An adaptive inverse-distance weighting spatial interpolation technique},
journal = {Computers \& Geosciences},
volume = {34},
number = {9},
pages = {1044-1055},
year = {2008},
issn = {0098-3004},
author = {George Y. Lu and David W. Wong},
}

@article{adhikary2013joint,
  title   = {Joint spatial division and multiplexing—The large-scale array regime},
  author  = {Adhikary, Ansuman and Nam, Junyoung and Ahn, Jae-Young and Caire, Giuseppe},
  journal = IEEE_J_IT,
  volume  = {59},
  number  = {10},
  pages   = {6441--6463},
  year    = {2013},
  month   = oct
}

@ARTICLE{9354041,
  author={Levie, Ron and Yapar, Cağkan and Kutyniok, Gitta and Caire, Giuseppe},
  journal={IEEE Trans. Wireless Commun.}, 
  title={RadioUNet: Fast Radio Map Estimation With Convolutional Neural Networks}, 
  year={2021},
  volume={20},
  number={6},
  pages={4001-4015}}

@ARTICLE{10130091,
  author={Zhang, Songyang and Wijesinghe, Achintha and Ding, Zhi},
  journal={IEEE Internet Things J.}, 
  title={RME-GAN: A Learning Framework for Radio Map Estimation Based on Conditional Generative Adversarial Network}, 
  year={2023},
  volume={10},
  number={20},
  pages={18016-18027},}

@ARTICLE{10843401,
  author={Luo, Xuanhao and Li, Zhizhen and Peng, Zhiyuan and Chen, Mingzhe and Liu, Yuchen},
  journal={IEEE Trans. Cognit. Commun. Networking}, 
  title={Denoising Diffusion Probabilistic Model for Radio Map Estimation in Generative Wireless Networks}, 
  year={2025},
  volume={11},
  number={2},
  pages={751-763}}

@article{wu2021digital,
  title   = {Digital twin networks: A survey},
  author  = {Wu, Yiwen and Zhang, Ke and Zhang, Yan},
  journal = IEEE_J_IOT,
  volume  = {8},
  number  = {18},
  pages   = {13789--13804},
  year    = {2021},
  month   = sep
}

@article{8226757,
  author  = {del Peral-Rosado, José A. and Raulefs, Ronald and López-Salcedo, José A. and Seco-Granados, Gonzalo},
  journal = IEEE_J_WC,
  title   = {Survey of Cellular Mobile Radio Localization Methods: From {1G} to {5G}},
  year    = {2018},
  month   = {2nd Quart.},
  volume  = {20},
  number  = {2},
  pages   = {1124-1148},
  doi     = {10.1109/COMST.2017.2785181}
}

@article{10081412,
  author  = {Croitoru, Florinel-Alin and Hondru, Vlad and Ionescu, Radu Tudor and Shah, Mubarak},
  journal = IEEE_J_PAMI,
  title   = {Diffusion Models in Vision: A Survey},
  year    = {2023},
  month   = sep,
  volume  = {45},
  number  = {9},
  pages   = {10850-10869},
  doi     = {10.1109/TPAMI.2023.3261988}
}

\vfill

\end{document}